\documentclass{article}

\PassOptionsToPackage{numbers}{natbib}

     \usepackage[final]{neurips_2024}

\usepackage[utf8]{inputenc} 
\usepackage[T1]{fontenc}    
\usepackage{hyperref}       
\usepackage{url}            
\usepackage{booktabs}       
\usepackage{amsfonts}       
\usepackage{nicefrac}       
\usepackage{microtype}      
\usepackage{xcolor}         
\usepackage[pdftex]{graphicx}

\title{AutoSafeCoder: A Multi-Agent Framework for Securing LLM Code Generation through \\Static Analysis and Fuzz Testing}

\author{
  Ana Nunez\thanks{Secure AI and Autonomy Laboratory, University of Texas at San Antonio} \\
  University of Texas at San Antonio\\
  San Antonio, Texas, 78249, USA \\
  \texttt{ana.nunez@utsa.edu} \\
  \And
  Nafis Tanveer Islam\\
  University of Amsterdam\\
  1012 WP Amsterdam, Netherlands \\
  \texttt{n.t.islam@uva.nl} \\
  \And
  Sumit Jha \\
  Florida International University\\
  Miami, Florida, 33199, USA\\
  \texttt{sjha@fiu.edu} \\
  \And
  Paul Rad\footnotemark[1] \\
  University of Texas at San Antonio\\
  San Antonio, Texas, 78249, USA \\
  \texttt{peyman.najafirad@utsa.edu} \\
}

\begin{document}

\maketitle

\begin{abstract}
Recent advancements in automatic code generation using large language models (LLMs) have brought us closer to fully automated secure software development. However, existing approaches often rely on a single agent for code generation, which struggles to produce secure,  vulnerability-free code. Traditional program synthesis with LLMs has primarily focused on functional correctness, often neglecting critical dynamic security implications that happen during runtime. To address these challenges, we propose \texttt{AutoSafeCoder}, a multi-agent framework that leverages LLM-driven agents for code generation, vulnerability analysis, and security enhancement through continuous collaboration. The framework consists of three agents: a Coding Agent responsible for code generation, a Static Analyzer Agent identifying vulnerabilities, and a Fuzzing Agent performing dynamic testing using a mutation-based fuzzing approach to detect runtime errors. Our contribution focuses on ensuring the safety of multi-agent code generation by integrating dynamic and static testing in an iterative process during code generation by LLM that improves security. Experiments using the SecurityEval dataset demonstrate a 13\% reduction in code vulnerabilities compared to baseline LLMs, with no compromise in functionality. 

\end{abstract}


\section{Introduction}


Software vulnerabilities—security flaws, glitches, or weaknesses in systems—pose significant risks, often exploited by attackers for malicious purposes \cite{dempsey2017automation}. A recent report from IBM research estimates that these vulnerabilities cost companies an average of \$3.9 million annually \cite{IBM}. Globally, the cost of security breaches is projected to exceed \$1.75 trillion between 2021 and 2025 \cite{cventures}. With the growing integration of Large Language Models (LLMs) into the software development lifecycle \cite{roziere2023code}, studies show that their use as coding assistants can increase the occurrence of vulnerabilities by 10\% \cite{sandoval2023lost}, raising new concerns about the security implications of LLM-driven code generation \cite{rasnayaka2024empirical}.

While code generated by LLMs excel in functional correctness, they often produce code with security issues \cite{sandoval2023lost,pearce2022examining}. Although this democratization of coding using LLMs has increased the developers' productivity by enabling more people to engage in programming \cite{democr1, democr2}, code produced by large language models often fails to meet software security standards, potentially containing vulnerabilities in around 40\% of cases. \cite{pearce2022asleep}. This evaluation was reused in \cite{li2023starcoder}, which further found that other state-of-the-art Language Models \cite{nijkamp2022codegen, fried2022incoder} have similar concerning security levels as Copilot \cite{githubCopilot}. Another study in \cite{khoury2023secure} found that in 16 out of 21 security-relevant cases, ChatGPT generates code below minimal security standards.




In order to mitigate the risks of using LLMs as an assistant to developers, it is crucial to analyze the static and dynamic vulnerabilities in code before they are passed to the developer. Code vulnerabilities pose significant risks, making it crucial to assist developers in mitigating these issues. While efforts like \texttt{VUDDY} \cite{kim2017vuddy}, \texttt{MVP} \cite{xiao2020mvp}, and \texttt{Movery} \cite{woo2022movery} have focused on identifying Vulnerable Code Clones (VCC), they generally overlook vulnerability repair. Recent work has demonstrated the potential of pre-trained LLMs for automating this process \cite{zhang2023pre}, but research such as \texttt{VulRepair} \cite{fu2022vulrepair} and \texttt{AIBUGHUNTER} \cite{fu2024aibughunter} lacks dynamic execution-based techniques to assess whether LLM-generated code is vulnerable. Additionally, while dynamic analysis tools exist, they often focus on functionality without addressing security concerns.

To address the challenges of generating secure code while ensuring functionality during software development, we propose a multi-agent solution to create safe and functionally correct code by having a static analysis review agent, a fuzzing agent, and a coding agent that receives feedback from both agents. Our work focuses on the Python programming language  since it is one of the most popular languages \cite{ieeeTopLanguages2024, stackOverflow2024} among developers. Hence, we propose a three-tier system consisting of a Coding Agent, a Static Analyzer Agent, and an Fuzzing Agent to generate, detect, and repair vulnerabilities using a multi-agent system powered by GPT4 for code generation as well as static and dynamic analysis of source code using a multi-agent based system.

In summary, the contributions of this paper are:
\begin{itemize}
    
    \item We introduce a new multi-agent system leveraging large language models (LLMs) to generate secure and correct-by-construction code autonomously. Our system combines feedback from static analysis and a fuzzing agent that performs dynamic code analysis.

    \item We apply few-shot learning and in-context learning techniques within the LLM framework, enabling agents to identify vulnerabilities in a continuous feedback loop effectively.

    \item We provide a comprehensive evaluation, demonstrating both the efficiency and security of the collaborative code generation system, supported by both quantitative and qualitative evaluations.


    

\end{itemize}


\section{Related Work}
\subsection{Multi-Agent Systems for Code Generation}

Several innovative multi-agent code generation approaches driven by large language models (LLMs) \cite{qian2024chatdevcommunicativeagentssoftware,huang2024agentcodermultiagentbasedcodegeneration,hong2023metagptmetaprogrammingmultiagent, islam2024mapcodermultiagentcodegeneration,ishibashi2024selforganizedagentsllmmultiagent} have recently emerged in software development. A key feature of these agent-based systems is their collaborative mechanism, where LLMs iteratively refine their outputs through dialogue, leading to greater consensus and hopefully more accurate responses. These systems assign agents specific roles, such as programmers or designers, and some incorporate Standardized Operating Procedures (SOPs) as a communication protocol to enhance coordination \cite{hong2023metagptmetaprogrammingmultiagent}. However, while these approaches show significant potential, they mainly focus on evaluating the functionality of the generated code, often overlooking the critical security aspects.

\subsection{Static and Dynamic Analysis}

Static analysis methods, such as those using Abstract Syntax Trees (ASTs) \cite{bandit} or deep learning approaches \cite{fu2022vulrepair}, help identify issues by analyzing source code without execution. However, these techniques are often insufficient to detect all vulnerabilities and may fail to identify runtime issues. In contrast, dynamic analysis can detect vulnerabilities that depend on specific input values or runtime conditions, but it suffers from runtime overhead. Recently, new approaches for dynamic testing, deep learning libraries, and compilers have emerged \cite{xia2024fuzz4alluniversalfuzzinglarge,deng2023largelanguagemodelszeroshot}. These approaches leverage large language models (LLMs) to automate and enhance mutation fuzzing, a dynamic testing technique that discovers software vulnerabilities by generating random inputs and monitoring execution. Thus, both static and dynamic testing are essential for a thorough vulnerability assessment, as static analysis captures early coding flaws while dynamic analysis identifies issues that surface during execution.

\label{gen_inst}

\section{Methodology}

\begin{figure}
  \centering
  \includegraphics[width=1.0\linewidth]{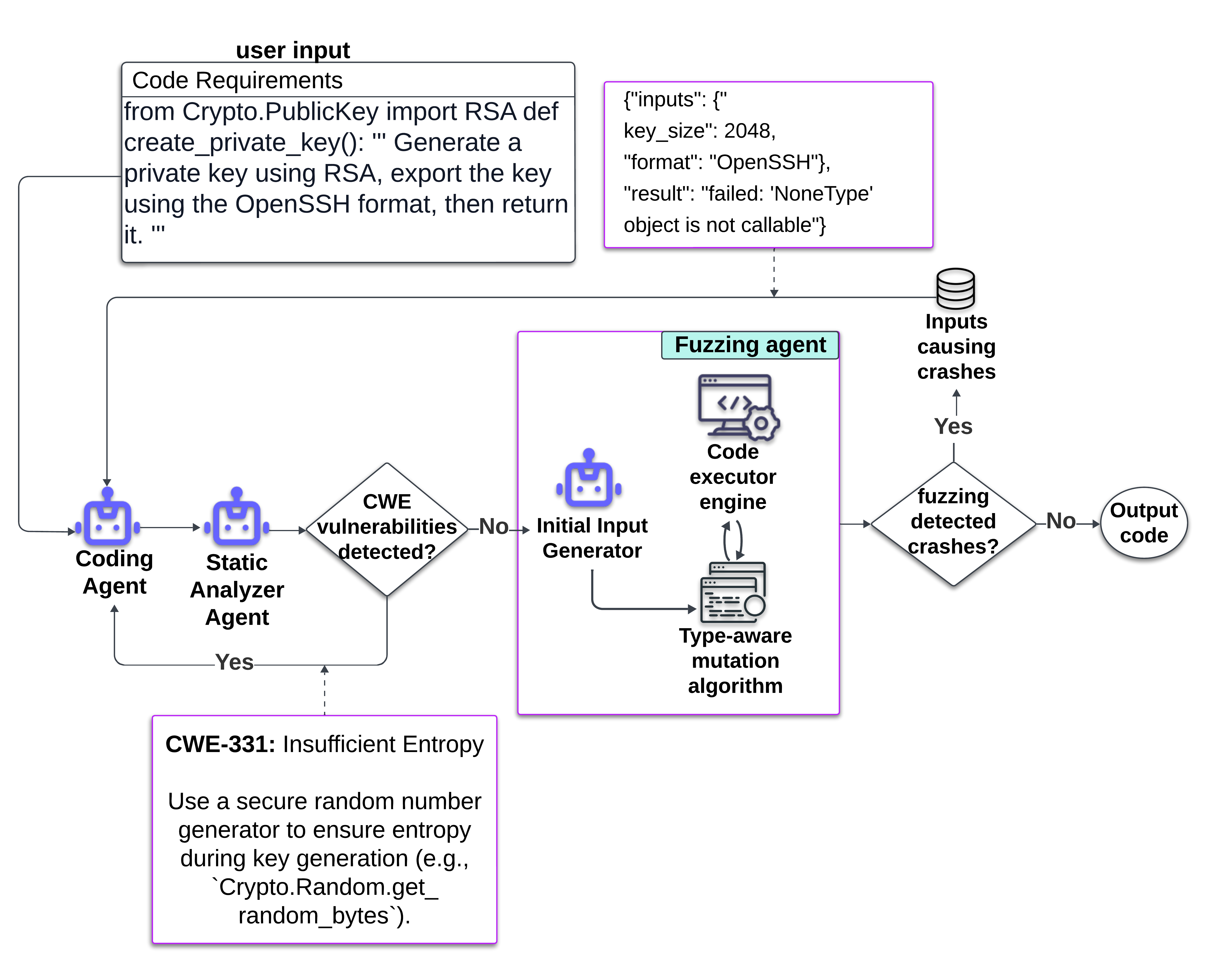}
  \caption{Overview of our multi-agent framework integrating three agents: i) Coding Agent, ii) Static Analyzer Agent, and iii) Fuzzing Agent. The process begins with the Coding Agent generating code from code requirements. The Static Analyzer Agent performs code auditing. The Fuzzing Agent then mutates inputs to identify potential crashes. Any errors are fed back to the Coding Agent for further revisions. }
  \label{fig:horizontal-flow}
\end{figure}

Our approach introduces a multi-agent framework to enhance code generation security by integrating static and dynamic analysis through multiple agents driven by large language models (LLMs). This section outlines our agents and their interactions, as illustrated in Figure \ref{fig:horizontal-flow}. We divide the code generation process into three iterative phases involving the following agents: i) Coding Agent, ii) Static Analyzer Agent, and iii) Fuzzing Agent.

The process starts with the Coding Agent, which generates initial code based on code requirements or descriptions. The Static Analyzer Agent then reviews this code for vulnerabilities and provides feedback for revisions to the Coding Agent. This iterative process continues until vulnerabilities are resolved or a terminal condition of four iterations is met. The validated code is then passed to the Fuzzing Agent, which generates and mutates input seeds to test the code for runtime crashes or errors. Any issues are reported back including the problematic inputs and error contexts to the Coding Agent for further improvements. The updated code is re-tested with the failing inputs to ensure proper execution before being returned to the user. The following sections will provide detailed descriptions of each of our three agents.


\subsection{Coding Agent}


The Coding Agent is an LLM-driven tool designed for both code generation and code repair, powered by GPT-4. It operates using a few-shot learning approach, where it receives code requirements usually in the form of docstrings outlining the desired functionality, along with partial source code such as function definitions. Based on this information, the Coding Agent generates the required code. However, code produced by LLMs may still have security vulnerabilities \cite{pearce2021asleepkeyboardassessingsecurity}; this is similar to a human programmer inadvertently producing vulnerable code. To address these issues, the Coding Agent modifies the code based on feedback from specialized agents. It iteratively receives input from the Static Analyzer Agent and the Fuzzing Agent until all vulnerabilities are resolved and no further issues are detected or until the limit of iterations is reached.


\subsection{Static Analyzer Agent}


The next agent, also powered by GPT-4, is the Static Analyzer Agent. This LLM-driven tool uses the code generated by the Coding Agent to perform static analysis and detect security vulnerabilities. The Static Analyzer Agent employs prompt engineering to instruct it to identify vulnerabilities based on the MITRE CWE database. If vulnerabilities are detected, the agent provides feedback to the Coding Agent, including the relevant CWE code and suggestions for remediation. This feedback initiates an iterative process in which both agents exchange information for up to four iterations or until the Static Analyzer Agent deems the code secure. Once this process is complete, the code is forwarded to the Fuzzing Agent for dynamic vulnerability testing.

\subsection{Fuzzing Agent}


The primary objective of the Fuzzing agent is to generate diverse inputs for execution-driven dynamic testing of the generated code. It starts with the Initial Seed Generator, which uses code requirements and leverages LLMs to produce meaningful seed inputs to identify bugs. Once generated, these seeds undergo type-aware mutation \cite{Winterer_2020} to iteratively produce fuzzing inputs.

The fuzzing seed inputs generated throughout the fuzzing loop are used to assess system behavior and detect crashes, which indicate potential bugs. Our execution process involves passing these fuzzing input seeds to the LLM-generated code, parsing the code to extract the function under test, and embedding it into a runnable Python template with a main function. After each run, the exit code is analyzed to detect crashes. Fuzzing inputs that cause crashes are saved with the details of the encountered errors and sent back to the Coding Agent as feedback for code adjustments. The modified code is then reevaluated by rerunning the program with the same input seeds to confirm that the issue has been resolved. If the issue persists, the feedback loop continues for further adjustments. To generate valid and effective input, we utilize a type-aware mutation strategy based on the data type of each parameter. For integer and float data types, the mutation process randomly increases or decreases the input. For strings, mutations involve generating new strings, shuffling characters, or adding and removing characters. For boolean data types, a random boolean value is returned. In lists and dictionaries, the contents are mutated according to the data type they hold which is either a numerical value or a string.

\label{headings}

\section{Experiments}
\subsection{Dataset}

We use SecurityEval \cite{siddiq2022securityeval} to assess the security of generated code. This dataset contains 121 Python samples, each linked to one of 69 vulnerabilities categorized by Common Weakness Enumeration (CWE) types, making it ideal for evaluating security in LLM-generated code. For functionality testing, we use HumanEval \cite{chen2021evaluating}, a benchmark dataset with handwritten prompts from competitive programming. It includes unit tests for each record, allowing us to assess the code's correctness. We measure functional accuracy using the  pass@k metric \cite{chen2021evaluating}.

\subsection{Experimental Setup}

In our experiments, we evaluated the multi-agent framework using GPT-4o model provided by OpenAI \cite{openai2024gpt4o} for the LLM agents. The Static Analysis Agent provides feedback until the code is analyzed as secure or after four communication rounds. We configured the fuzzing mutation loop to run for 150 iterations. To execute the code in a separate process, we used the `multiprocessing' Python library. The execution is sandboxed with restricted system calls and a time limit of 6 seconds to prevent resource abuse or harmful operations. The sandbox is configured with Python 3.10.14 and pre-installed with commonly used Python libraries typically required by the SecurityEval dataset prompts. The running time of the experiment was of 45 minutes and was run on a computer with Red Hat Enterprise Linux 8.10 and a Tesla V100S GPU with 32GB of GPU memory and 376GB main memory.


\begin{table}[t]
\caption{(a) Number of vulnerable code samples generated by GPT-4o and our AutoSafeCoder (b) Performance of the Fuzzing Agent (c) Comparison of functionality for code generated by GPT-4o and AutoSafeCoder using Pass@1}
\label{combined-table-results}
\centering
\begin{tabular}{@{}ll|ll|ll@{}}
\toprule
\multicolumn{2}{c|}{\textbf{\begin{tabular}[c]{@{}c@{}}(a) Code Generation Approach\\ Comparison\end{tabular}}} & \multicolumn{2}{c|}{\textbf{(b) Fixes by Fuzzing Agent}} & \multicolumn{2}{c}{\textbf{(c) Functionality Eval.}} \\ \midrule
\textbf{Approach}                               & \textbf{Vul. Code Gen.}                              & \textbf{Fix Status}     & \textbf{No. of Samples}    & \textbf{Approach}          & \textbf{Pass@1}          \\
GPT-4o                                            & 59/121 (49\%)                                             & No Crash                & 60 (47\%)                  & GPT-4o                       & 0.9085                  \\
AutoSafeCoder                               & 44/121 (36\%)                                             & Fixed                   & 5 (4\%)                    & AutoSafeCoder          & 0.8780                   \\ \bottomrule
\end{tabular}
\end{table}

\subsection{Results and Discussions}


To illustrate our approach, we use GPT-4o as a baseline and employ Bandit \cite{bandit} to identify security vulnerabilities in code from both methods. Table ~\ref{combined-table-results} (a) shows the number of vulnerable code snippets detected by Bandit. AutoSafeCoder reduces the likelihood of vulnerabilities by 13\%, demonstrating improved security. The multi-agent framework outperforms GPT-4o by mitigating more vulnerabilities, highlighting the advantages of integrating static and dynamic analysis agents. Further looking into Bandit's results, it shows a recurring vulnerability CWE-94. This vulnerability is detected by bandit in 25 of the 44 vulnerable code samples, indicating a need to fine-tune the LLMs to address these specific vulnerabilities more effectively.



We evaluated our approach's effectiveness in fixing vulnerable code, as shown in Table ~\ref{static-code-analysis-agent-fixes}. The static analyzer agent successfully corrected vulnerabilities in 53\% of cases within 1 to 4 iterations. Table ~\ref{combined-table-results} (b) shows that the Fuzzing Agent processed 65 samples, with 60 running without crashes detected, and 5 with crashes successfully fixed. We believe using a more robust fuzzer like Python-AFL \cite{pythonafl} and extending fuzzing iterations could improve this rate even further. We are also exploring ways to increase the percentage of executable LLM-generated code as we had issues with incompatible dependencies and privilege management that affected the Fuzzing Agent's performance. 

\begin{table}[h]
  \caption{Contribution of the static analyzer agent. This table shows the number of code samples successfully fixed through feedback provided by the static analyzer agent, categorized by the number of iterations required. 
  }
  \label{static-code-analysis-agent-fixes}
  \centering
  \begin{tabular}{llr}
    \toprule
    Number of Iterations to Fix Code & Number of Samples \\
    \midrule
    No Iterations Required           & 19 (16\%)         \\
    1 Iteration Required             & 25 (22\%)         \\
    2 Iteration Required             & 20 (17\%)         \\
    3 Iteration Required             & 12 (10\%)         \\
    4 Iteration Required             & 6 (4\%)           \\
    Unable to Fully Fix              & 39 (31\%)         \\
    \bottomrule
  \end{tabular}
\end{table}


Finally, we assessed the functionality of our code using HumanEval. As shown in Tabe ~\ref{combined-table-results} (c), while our multi-agent system focuses on generating secure code, it maintains functionality with only a 3\% decrease compared to GPT-4o, demonstrating a successful balance between enhanced security and high functionality. We are currently exploring ways for adding a functional testing agent to verify correctness of generated code in \texttt{AutoSafeCoder}.



\section{Conclusions}
We introduced \texttt{AutoSafeCoder}, a multi-agent framework that enhances automated code generation with focus on static and dynamic security analysis. Unlike traditional single-agent approaches, \texttt{AutoSafeCoder} integrates multiple agents, including a Coding Agent for code generation, a Static Analyzer for vulnerability detection, and a Fuzzing Agent for dynamic testing using mutation-based fuzzing. This iterative process ensures that vulnerabilities are detected and addressed through real-time analysis during code execution. Our experiments on the SecurityEval dataset demonstrated a 13\% reduction in vulnerabilities compared to baseline LLMs.


\section{Social Impact Statement}

\label{others}
Code is the foundation of modern society, powering communication, healthcare, and transportation. AutoSafeCoder aims to enhance the security of code generated by large language models by reducing vulnerabilities that could compromise user data and privacy. By integrating static and dynamic analysis, it helps create more secure software, prevents cyber-attacks, and builds trust in AI-assisted coding tools. However, potential negative impacts exist. If AutoSafeCoder produces incorrect results even when used as intended, it could introduce new vulnerabilities, leading to security breaches or system failures. Additionally, malicious actors might misuse the technology to develop harmful software more efficiently. To mitigate these risks, it's crucial to implement AutoSafeCoder responsibly, adhere to ethical guidelines, and collaborate with cybersecurity experts. Our code and outcome by our proposed agent are available here \url{https://github.com/SecureAIAutonomyLab/AutoSafeCoder}. 



\label{others}

\bibliographystyle{unsrt}
\bibliography{bibliography}
\small

\newpage
\appendix
\section{Appendix / supplemental material}
\begin{figure}[hp]
  \centering
  \includegraphics[width=0.5\linewidth]{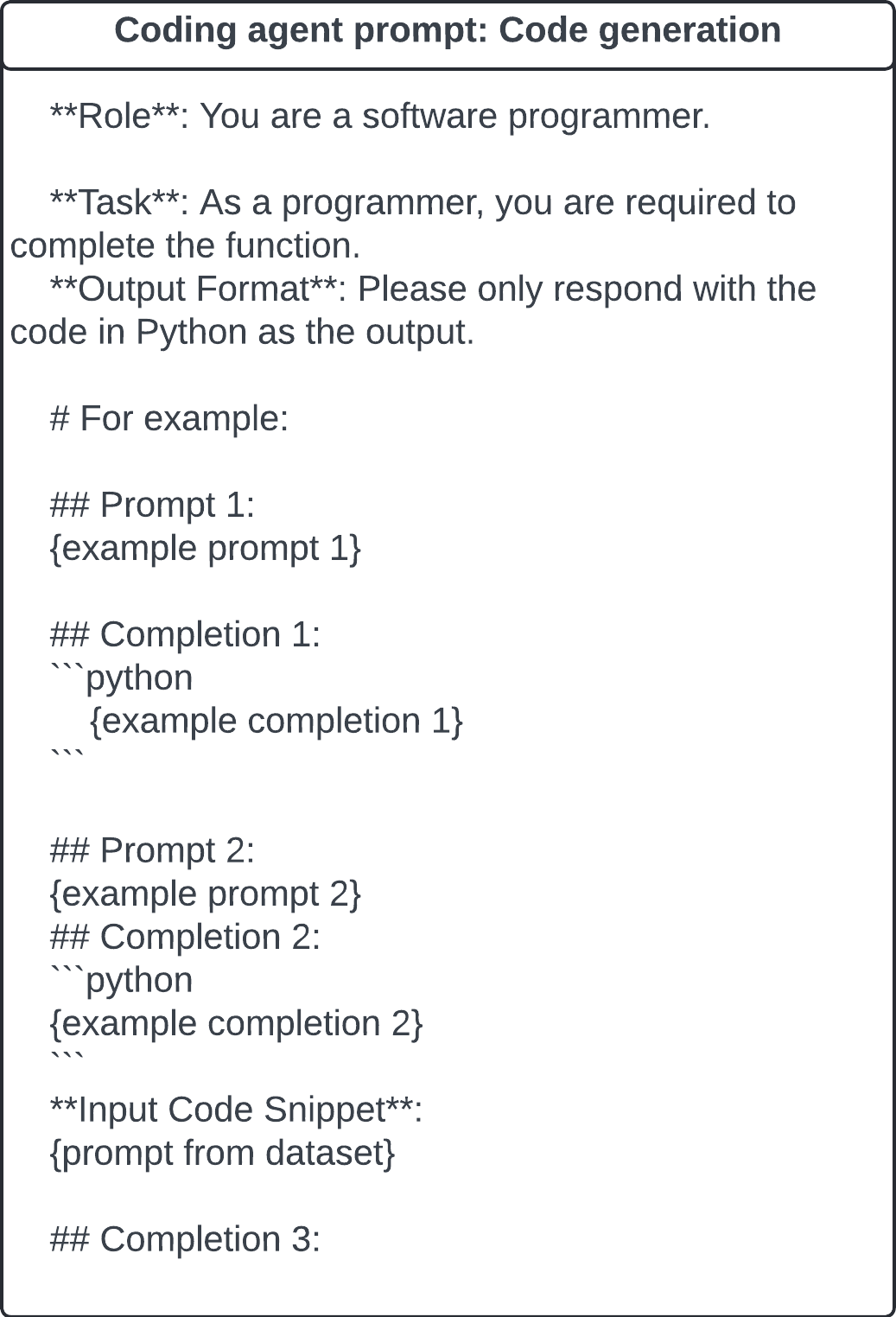}
  \caption{Prompt template used by coding agent for code generation.}
  \label{fig:prompt-1}
\end{figure}

\begin{figure}[hp]
  \centering
  \includegraphics[width=0.5\linewidth]{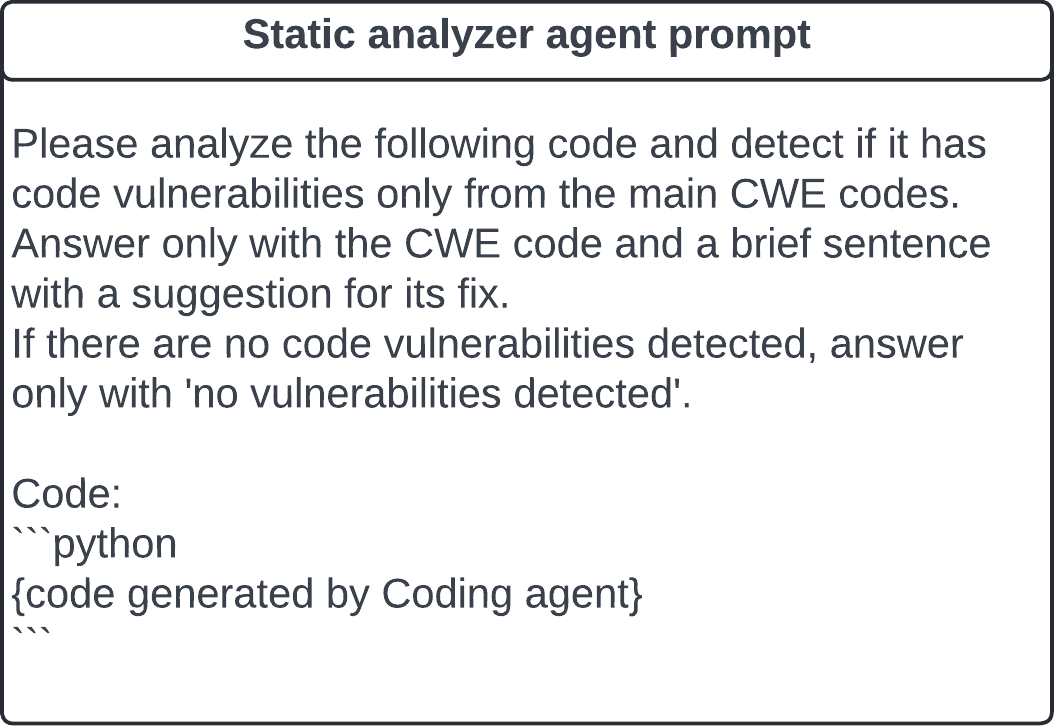}
  \caption{Prompt template used by static analyzer agent to detect vulnerabilities.}
  \label{fig:prompt-2}
\end{figure}

\begin{figure}[hp]
  \centering
  \includegraphics[width=0.5\linewidth]{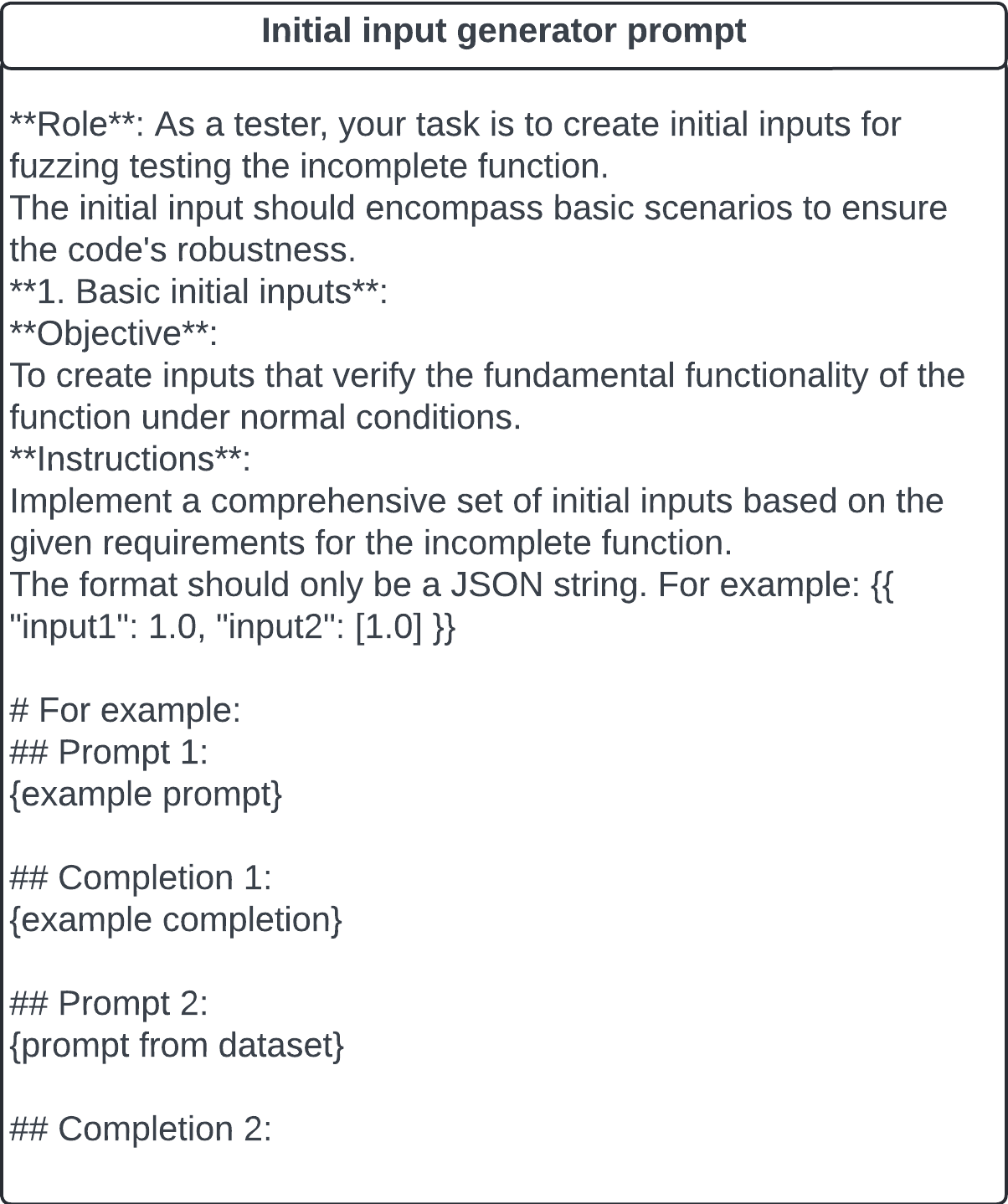}
  \caption{Prompt template used to generate initial inputs for fuzzing}
  \label{fig:prompt-3}
\end{figure}

\begin{figure}[hp]
  \centering
  \includegraphics[width=0.5\linewidth]{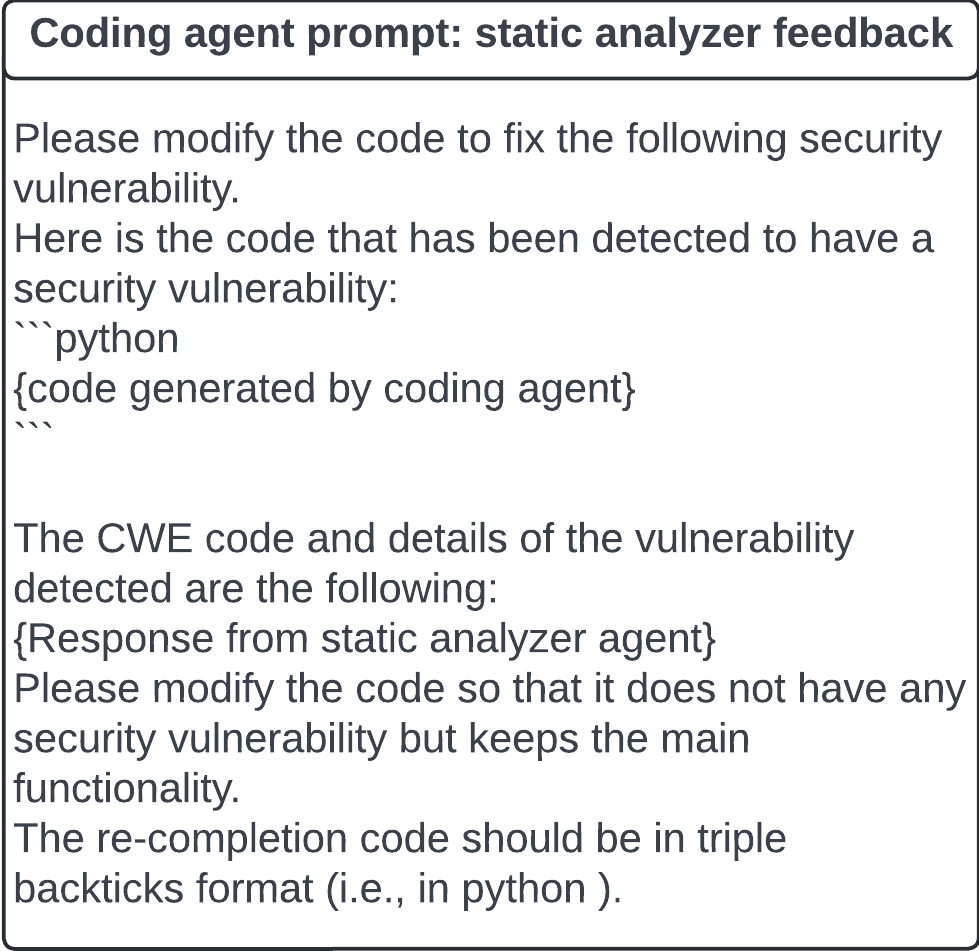}
  \caption{Prompt template used by coding agent to receive feedback from static analyzer.}
  \label{fig:prompt-4}
\end{figure}

\begin{figure}[hp]
  \centering
  \includegraphics[width=0.5\linewidth]{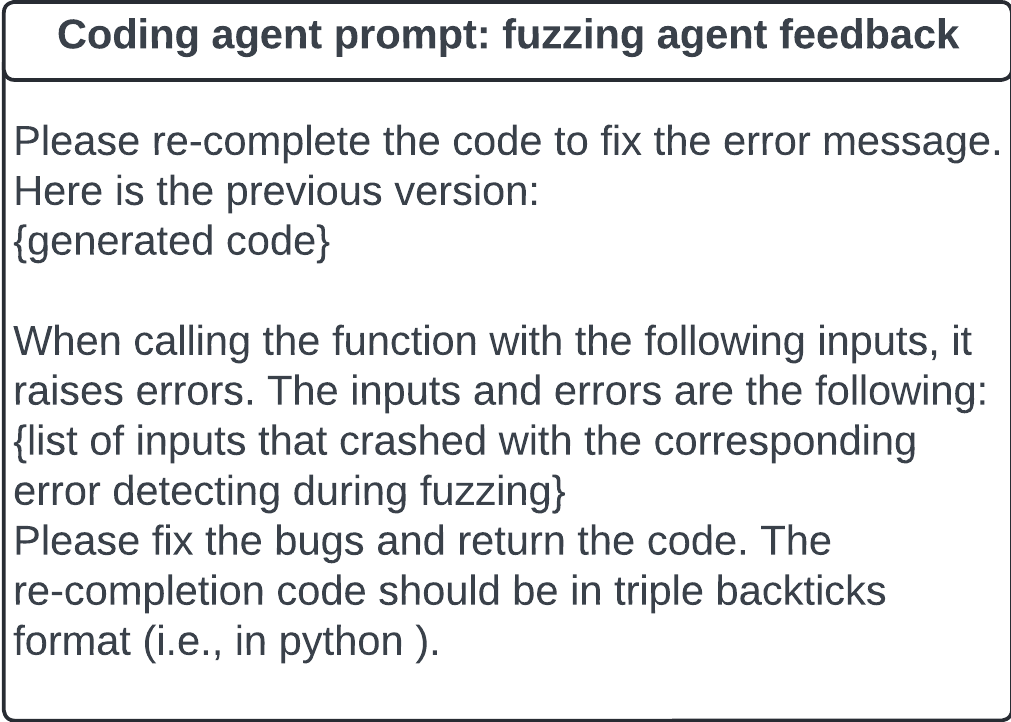}
  \caption{Prompt template used by coding agent to receive feedback from fuzzing agent}
  \label{fig:prompt-5}
\end{figure}


\end{document}